\documentclass{jpp}

\usepackage{graphicx}
\usepackage[authoryear]{natbib}

\usepackage[update,prepend]{epstopdf}
\usepackage{cite}
\usepackage{url}
\usepackage[colorlinks=true,
            linkcolor=red,
            citecolor=blue,
            urlcolor=blue]{hyperref}
\usepackage[normalem]{ulem}

\newcommand{\eps}{\epsilon}
\newcommand{\epsL}{\epsilon_{\rm L}}
\newcommand{\epsR}{\epsilon_{\rm R}}
\newcommand{\epsS}{\epsilon_{\rm S}}
\newcommand{\npar}{n_{\|}}
\newcommand{\nperp}{n_{\perp}}
\newcommand{\om}{\omega}
\newcommand{\ompi}{\omega_{pi}}
\newcommand{\omci}{\omega_{ci}}

\newcommand{\zn}{\mathcal{Z}}
\newcommand{\kpar}{k_{\|}}
\newcommand{\Na}{\rm{Na}}
\newcommand{\beq}{\begin{equation}}
\newcommand{\eeq}{\end{equation}}

\ifCUPmtlplainloaded \else
  \checkfont{eurm10}
  \iffontfound
    \IfFileExists{upmath.sty}
      {\typeout{^^JFound AMS Euler Roman fonts on the system,
                   using the 'upmath' package.^^J}%
       \usepackage{upmath}}
      {\typeout{^^JFound AMS Euler Roman fonts on the system, but you
                   dont seem to have the}%
       \typeout{'upmath' package installed. JPP.cls can take advantage
                 of these fonts, if you use 'upmath' package.^^J}%
      }
  \else
  \fi
\fi

\ifCUPmtlplainloaded \else
  \checkfont{msam10}
  \iffontfound
    \IfFileExists{amssymb.sty}
      {\typeout{^^JFound AMS Symbol fonts on the system, using the
                'amssymb' package.^^J}%
       \usepackage{amssymb}%
         \let\leq=\leqslant
         \let\geq=\geqslant
      }{}
  \fi
\fi

\ifCUPmtlplainloaded \else
  \IfFileExists{amsbsy.sty}
    {\typeout{^^JFound the 'amsbsy' package on the system, using it.^^J}%
     \usepackage{amsbsy}}
    {\providecommand\boldsymbol[1]{\mbox{\boldmath $##1$}}}
\fi

\newsavebox{\astrutbox}
\sbox{\astrutbox}{\rule[-5pt]{0pt}{20pt}}

\title[Tunneling and mode conversion in magnetospheric plasmas]
      {Tunneling and mode conversion of fast magnetosonic waves in the
        magnetospheres of Earth and Mercury}

\author[Ye.O. Kazakov and T. F\"{u}l\"{o}p]%
{Y\ls E\ls V\ls G\ls E\ls N\ns  O.\ns K\ls A\ls Z\ls A\ls K\ls O\ls V$^1$%
  \thanks{Email address for correspondence: yevgen.kazakov@rma.ac.be}\ns
\and T\ls \"{U}\ls N\ls D\ls E\ns F\ls \"{U}\ls L\ls \"{O}\ls P$^2$}

\affiliation{$^1$Laboratory for Plasma Physics, Royal Military Academy, LPP-ERM/KMS, TEC Partner,
Brussels, BE-1000, Belgium \\[\affilskip]
$^2$Department of Applied Physics, Chalmers University of Technology, G\"{o}teborg, SE-41296, Sweden}

\pubyear{2010}
\volume{650}
\pagerange{119--126}
\date{?; revised ?; accepted ?. - To be entered by editorial office}
\begin{document}

\maketitle

%%%%%%%%%%%%%%%%%%%%%%%%%%%%%%%%%%%%%%%%%%%%%%%%%%%%%%%%%%%%%%%%%%%%%%%%%%
%   Abstract:
%%%%%%%%%%%%%%%%%%%%%%%%%%%%%%%%%%%%%%%%%%%%%%%%%%%%%%%%%%%%%%%%%%%%%%%%%%

\begin{abstract}
Narrow-band linearly polarized waves, having a~resonant structure and
a peak frequency between the local cyclotron frequency of protons and
heavy ions, have been detected in the magnetospheres of Earth and of
Mercury. Some of these wave events have been suggested to be
driven by linear mode conversion (MC) of the fast magnetosonic waves
at the ion-ion hybrid (IIH) resonances.  Since the resonant IIH
frequency is linked to the plasma composition, solving the inverse
problem allows one to infer the concentration of the heavy ions from
the measured frequency spectra. In~this paper, we identify the
conditions when the MC efficiency is maximized in the magnetospheric
plasmas and discuss how this can be applied for estimating the heavy
ion concentration in the magnetospheres of Earth and Mercury.
\end{abstract}

\begin{PACS}
%Authors should not enter PACS codes directly on the manuscript, as these must be chosen during the online submission process and will then be added during the typesetting %process (see http://www.aip.org/pacs/ for the full list of PACS codes)
\end{PACS}

%%%%%%%%%%%%%%%%%%%%%%%%%%%%%%%%%%%%%%%%%%%%%%%%%%%%%%%%%%%%%%%%%%%%%%%%%%
%   Paper starts here: introduction
%%%%%%%%%%%%%%%%%%%%%%%%%%%%%%%%%%%%%%%%%%%%%%%%%%%%%%%%%%%%%%%%%%%%%%%%%%

\section{Introduction}

Heavy ion species are typically present in the magnetospheric plasmas
surrounding the planets. The origin of heavy ions can be quite
different depending on the planetary environment. For example, a
significant fraction of $\rm{He}^{+}$ and $\rm{O}^{+}$ ions,
originating from the polar wind and the outflow of ionospheric
ions is present in some regions of the terrestrial
magnetosphere~\citep{andre, yau}.  The magnetosphere of Mercury is
known to have a powerful exospheric source of $\rm{Na}^{+}$ ions due
to the sputtering of the planet's surface and subsequent
photoionization.  \citet{ip} estimated the content of the sodium ions
to be $\geq 10\%$, and \citet{cheng} concluded that the sodium ion
source is at least comparable to, or can be greater than the solar
wind source of protons.  Recent MESSENGER measurements also confirmed
that in some regions near Mercury, especially the nightside equatorial
region, the ${\rm Na}^{+}$ pressure can be a substantial fraction of
the proton pressure~\citep{zurbuchen}.  Another interesting example of
the heavy ion abundance is the inner Jovian magnetosphere.  It is
contaminated with sulfur and oxygen ions due to the intense volcanic
activity of one of Jupiter's moon Io, spewing plumes of sulfur
dioxide~\citep{russell:io, bagenal}.

Plasmas support multiple types of waves and instabilities, and the
resonant wave-particle interaction is possible in several frequency
domains.  Waves in the ion cyclotron (IC) range of frequencies are
particularly sensitive to the presence of various ion species in the
plasma. Resonant wave events in the IC frequency range
are known to be strongly dependent on the plasma ion
composition. Furthermore, the presence of multiple ion species, even
with small concentrations, can lead to the appearance of new and
modified resonance, cutoff, and crossover frequencies~\citep{andre:1}.
The origin of the waves can be different, and
various types of polarization (primarily left-hand, right-hand, and linearly polarized waves)
have been reported in the literature~(e.g. \citet{young, boardsen}).

Particle instruments have difficulties measuring the properties
  of the cold plasma because of spacecraft charging effects and also
  simply because the flux of low velocity particles is low. As
  mentioned in \citet{denton}, despite many years of magnetospheric
  research, there exists little precise knowledge of the bulk plasma
  composition within the inner magnetosphere. This is not only due to
  measurement difficulties but also, because the plasma composition is
  changing in space and time. \citet{denton} reported  major changes
  in the bulk plasma composition at geosynchronous orbit
  with respect to the solar cycle. Also, \citet{craven} highlighted,
  that the big spread in the data on He$^+$/H$^+$ density
  ratio reflects its significant dependence on the latitude and
  local magnetic time.

Already in one of the first papers on wave propagation in multi-ion
component plasmas, ~\citet{smith} noted that the ion-ion hybrid (IIH)
resonance and cutoff frequencies present in plasmas with two or more
ion species carry the information on ion densities. This motivated the
development of indirect methods to determine the plasma composition
using wave polarization analysis. Such an approach was used by
\citet{stasiewicz}, who inferred the plasma composition using the
lower hybrid and the IIH frequencies observed by the Freja satellite.
They estimated the concentrations of H$^{+}$, He$^{+}$ and O$^{+}$
ions to be 23\%, 19\% and 58\%, respectively, at an altitude of
1700~km above the Earth. Later, \citet{takahashi} analyzed ion
  composition in the plasma trough and plasma plume for the CRRES
  spacecraft orbit 962.  They reported significant concentrations of
  oxygen ions in the low density region of plasma trough $(L \simeq
  3.8-5.7)$: ${\rm O}^{+}$ ions were found to carry $\sim 50\%$ of the
  number density and $\sim 90\%$ of the mass density.  However, oxygen
  was shown to be much less abundant in the plasma plume $(L \geq 6)$.

Polarization analysis was recently applied by \citet{sakaguchi}
  to compute the ion composition in the deep inner magnetosphere using
  the Akebono satellite measurements. Negligible oxygen content was
  reported and the ion composition (H$^+$,He$^+$,O$^+$)
  =(83\%,16\%,1\%) was retrieved from the measured wave spectrogram.
  Relying on GOES-8 and GOES-10 data, somewhat lower helium
  concentrations $(X[{\rm He^{+}}]=6-16\%)$ were inferred by
  \citet{fraser1} for the plasma near the geostationary orbit.
  According to \citet{fraser2}, ion composition measurements
  identified He$^+$ as the second most abundant ion in the
  plasmasphere after protons.  However, the oxygen content was
  observed to enrich during substorms and just inside the plasmapause
  (``oxygen torus''), where its density was found to increase by a
  factor of 10 whereas there is no corresponding variation in helium
  and proton concentrations.

Analysis of the wave polarization can give insight into
  the physical characteristics of waves and help understand their
  propagation and excitation mechanisms.  In~this paper, we focus on
studying a particular type of the ultra-low-frequency (ULF) waves that
are linearly polarized within a narrow frequency band with
a resonant frequency below the local proton gyrofrequency, but larger
than the gyrofrequency of heavy ions.  Such IC waves have been
detected in the magnetospheres of Earth and of Mercury~\citep{young,
  perraut, russell, boardsen}.  There is an ongoing discussion on the
origin of these resonant waves. First, \citet{russell} argued that the
observed ULF wave was a~standing Alfv\'en wave along the magnetic
field line.  Then, \citet{othmer} developed a~model for the field line
resonances in Mercury's magnetosphere accounting for multi-ion plasma
composition and suggested the crossover frequency to be the preferred
resonant frequency.  Whereas for parallel propagation linear
  polarization occurs at the crossover frequency, fast magnetosonic
  (compressional) waves propagating across the magnetic field lines
  are polarized linearly at the IIH resonant frequency.
Alternatively, \citet{kim2008} performed wave simulations in
electron-proton-sodium plasmas and concluded that linearly polarized
IC waves can be generated due to mode conversion (MC), which the
fast wave undergoes at the the IIH and/or Alfv\'en resonances. The
additional IIH resonance appears in plasmas, which include at least
two ion species with the different charge-to-mass ratios, and gets its
name because the resonant frequency occurs between the two IC
frequencies.  Further numerical studies by \citet{kim2011} revealed
that wave absorption is much more efficient at the IIH resonance than
at the Alfv\'en resonance. In our recent papers \citep{prl2013,
  eps2014}, we complemented previous approaches and proved that
efficient tunneling and mode conversion requires the resonant IIH
frequency to be close to the crossover frequency (but somewhat
  below) for typical conditions of the magnetospheres of Earth and
Mercury. This means that the above mentioned explanations are not
contradictory to each other.

Since the resonant IIH frequency is linked to the plasma
composition, solving the inverse problem allows one to infer the
concentration of the heavy ions from the measured frequency
spectra. In~this paper, we identify numerically the conditions, when
the MC efficiency is maximized in the magnetospheric plasmas of Earth
and Mercury, and confirm analytical results derived
in~\citet{prl2013}. Whereas for Mercury's plasma, the resonant IIH
frequency is computed to be $\geq 0.8$ of the crossover frequency, for
the magnetospheric plasmas close to the geostationary orbit the IIH frequency is proved to be very
close to the crossover frequency. Due to arguments given above,
we will start our analysis considering two-ion proton-helium plasma at this region
and introduce later the effect of plasma contamination with oxygen ions.
This potentially allows to estimate the helium concentration from the recorded ULF spectrogram
relying on the two-ion plasma approximation or to provide an equation linking
He$^{+}$ and O$^{+}$ concentrations if the oxygen content can not be neglected.

The rest of the paper is organized as follows. In section~\ref{sect:2}
we discuss the conditions when MC is efficient in magnetospheric
plasmas. In section~\ref{sect:3} we supplement previous analytical
analysis with numerical MC computations and determine quantitatively
how close the resonant IIH frequency is to the crossover frequency for
the typical parameters of Earth's and Mercury's plasmas.  In
section~\ref{sect:4} we analyze the properties of the waves close to
the crossover condition and evaluate the range of the parallel wave
numbers yielding efficient MC.  Finally, in section~\ref{sect:5} we
evaluate the frequency width of the spectrum for Mercury's plasma and
compare it to the experimentally reported value in~\citet{russell}.
We~conclude our results in section~\ref{sect:6}.

\section{Conditions for efficient mode conversion}
\label{sect:2}

Propagation of fast magnetosonic waves (FW) through the inhomogeneous
magnetosphere is treated by using a 1D slab plasma
model~\citep{kim2008, kim2011, prl2013}.  In spite of its simplicity,
such a~model captures the main features of the resonant MC process. We
consider the incoming FW to be launched at the outer magnetosphere
side and study its penetration to the inner regions of the
magnetosphere.  The dispersion relation for FW propagating
predominantly across the magnetic field lines is approximately given
by~\citep{stix}
\beq n_{\perp, {\rm FW}}^2 =
\frac{(\epsL-\npar^2)(\epsR-n_{\|}^2)}{\epsS-n_{\|}^2},
\label{eq:fw.disp}
\eeq
where $\npar = c \kpar / \omega$ is the parallel (along the
magnetic field) refractive index, and $f=\omega/2\pi$ is the wave
frequency.  In Eq.~(\ref{eq:fw.disp}), $\epsS = \eps_1$, $\epsL =
\eps_1 - \eps_2$ and $\epsR = \eps_1 + \eps_ 2$ are the plasma
dielectric tensor components in the notation of Stix. In the cold-plasma
limit and for the IC frequency range the tensor components can be
written as
\beq
\eps_1 \simeq - \sum_{i} \frac{\ompi^2}{\om^2 - \omci^2}, \quad
\eps_2 \simeq - \sum_{i} \frac{\om}{\omci} \frac{\ompi^2}{\om^2 - \omci^2}, \quad
\eps_{\rm L,R} \simeq  \sum_{i} \frac{\ompi^2}{\omci(\omci \mp \om)},
\label{eq:tensor.comps}
\eeq
where the summation is to be taken over all ion species constituting
the plasma (protons and heavy ions), and $\omega_{pi}$ and
$\omega_{ci}$ are the plasma and cyclotron frequencies of ions,
respectively.  The vacuum and electron contributions to $\eps_{1}$
have been neglected because of their smallness with respect to the ion
terms.  Non-negligible electron contribution to $\epsilon_2$ has
already been included using the quasi-neutrality identity $\sum_{i}
\omega_{pi}^2/\omega_{ci} = \omega_{pe}^2/\left|\omega_{ce} \right|$.

\begin{figure}
\centering
\includegraphics[trim=0.cm 0cm -0cm -0mm, clip=true,
height=0.37\textwidth]{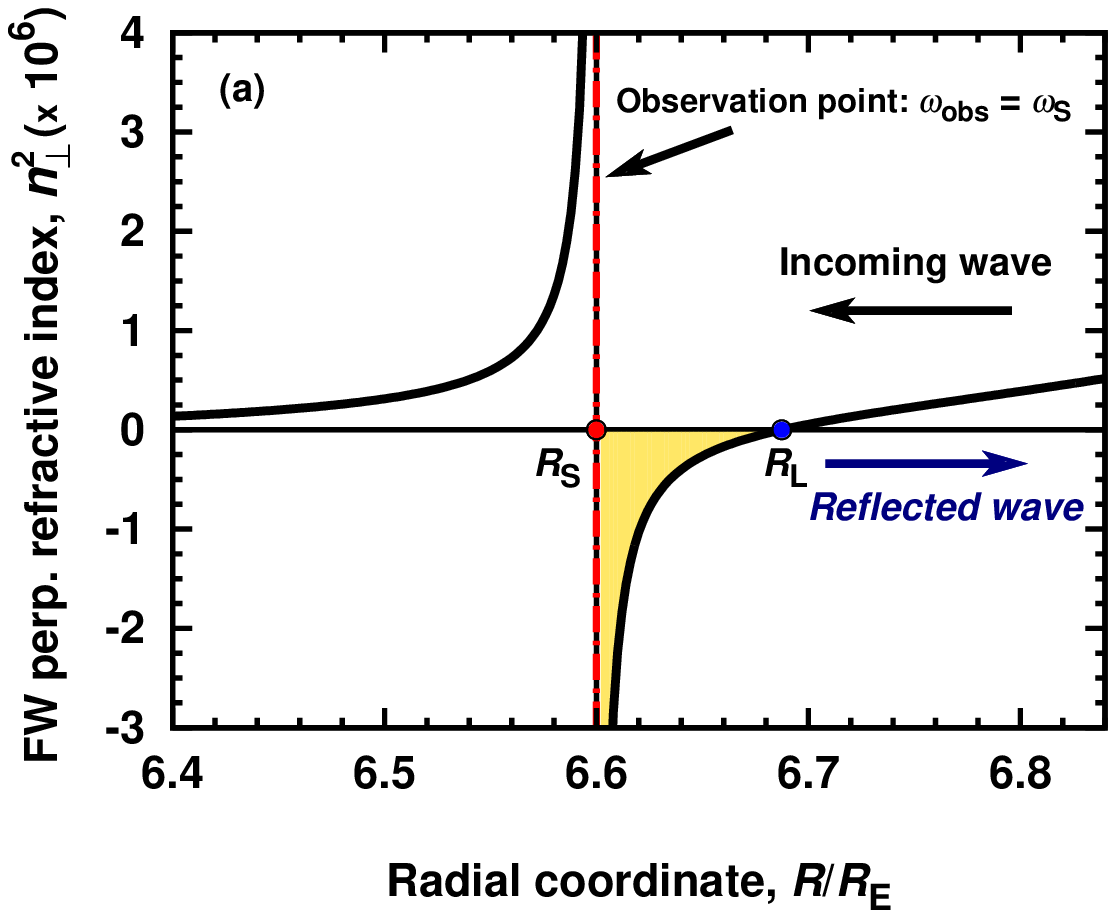}
%\hfill
\includegraphics[trim=0cm 0cm -0mm -0.0cm, clip=true,
height=0.37\textwidth]{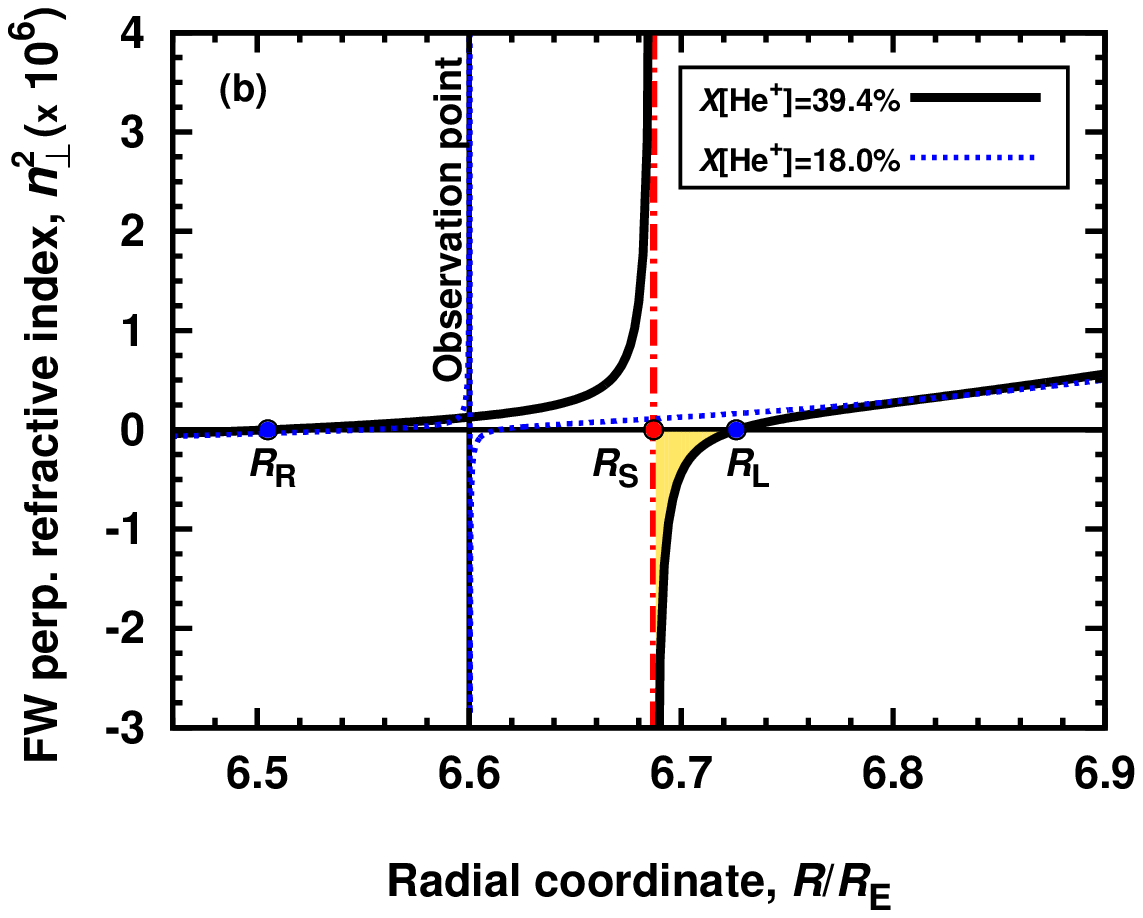}
\vspace{-2mm}
\caption{
Radial dependence of the FW dispersion $n_{\perp, {\rm FW}}^2$:
$f = 0.72~{\rm Hz}$,
$B(R)=B_0(R_{E}/R)^3$,  $n_{e}=10~{\rm cm}^{-3}$:
(a) $\kpar = 0$ and $X[{\rm He}^{+}]=39.4\%$;
(b) $\kpar = 5 \times 10^{-6}~{\rm m}^{-1}$ and $X[{\rm He}^{+}]=39.4\%/$
$18.0\%$ (solid/dotted line). Mode conversion layer is depicted as the shaded area.
\label{fig1}
}
\end{figure}

Figure~\ref{fig1} shows the radial dependence of $n_{\perp, {\rm FW}}^2$ (Eq.~(\ref{eq:fw.disp}))
computed for the plasma conditions
near the geostationary orbit.
The magnetospheric magnetic field
is intrinsically inhomogeneous and is assumed to vary
as $B(R)=B_{E}(R_{E}/R)^3$, where $B_{E}=31~{\rm \mu T}$ is the
magnetic field at magnetic equator of the Earth. The plots are computed
for the wave frequency $f=0.72~{\rm Hz}$ and plasma density $n_{e}=10~{\rm cm}^{-3}$,
and assuming plasma to be composed of protons and helium ions with
$X[{\rm He}^{+}]=n_{{\rm He}^{+}}/n_{e} = 39.4\%$.
Whereas Fig.~\ref{fig1}(a) stands for the special case of strictly perpendicular propagation $(\kpar = 0)$,
Fig.~\ref{fig1}(b) is plotted accounting for a finite $\kpar = 5 \times 10^{-6}~{\rm m}^{-1}$.
The radial variation of $\nperp^2$ depicted in Fig.~\ref{fig1} is typical for
plasmas composed of two ion species: one can clearly see
that there is a~radial region where $\nperp^2 \rightarrow \infty$.
It occurs when the wave frequency matches locally the ion-ion hybrid
frequency, which is defined by the condition
\begin{equation}
\epsS=\npar^2.
\label{eq:IIH}
\end{equation}
\noindent The IIH resonance does not appear in single-ion species plasmas and
arises if at least two ion species with different charge-to-mass ratios are present in a plasma.
The resonant IIH frequency
satisfies $\omega_{\rm c2} < \omega_{\rm S} < \omega_{\rm cH}$
(subindex `2' refers to heavy ions).
At the IIH resonant layer $R = R_{\rm S}$, the incoming FW is partially converted to a shorter wavelength mode
(slow/shear Alfv\'en wave or ion Bernstein wave).
Within the adopted model, this appears as the resonant FW absorption.
However, due to the radial inhomogeneity of the magnetic field
the IIH resonance layer
is accompanied by the L-cutoff layer $R=R_{\rm L}$ to the lower magnetic field side,
which is determined by the condition  $\epsL=\npar^2$.
Before reaching the resonance, the incoming FW first meets the cutoff
layer and is partially reflected there.  The IIH resonance and
L-cutoff layers form together the MC layer $(R_{\rm S} \leq R \leq R_{\rm
  L})$, which is a barrier for the propagating FW since within this
region the FW is evanescent, i.e. $n_{\rm \perp,\, FW}^2 < 0$.
The FW power transmits its power through the
MC layer via tunneling.  Only a fraction $\mathcal{T}=e^{-\pi \eta}$
of the incoming power tunnels through the layer, where $\eta$ is the
tunneling factor, roughly being a~product of the asymptotic FW
perpendicular wave number and the MC layer width. More rigorously, it
has been defined as \citep{swanson, kazakov2010}
\begin{equation}
\eta = \frac{2}{\pi} \int_{R_{\rm S}}^{R_{\rm L}} \left| k_{\rm \perp,FW}(R)\right| dR,
\label{eq:eta}
\end{equation}
where $k_{\perp, {\rm FW}}=(\om/c)n_{\perp, {\rm FW}}$ is the FW perpendicular wave number,
which is an imaginary number within the MC layer.

In case of low $\kpar$ and the isolated MC layer (Fig.~\ref{fig1}(a)), the reflection coefficient for the waves
excited at the outer magnetosphere (coming from the L-cutoff side) is given by
$\mathcal{R}=(1-\mathcal{T})^2$,
limiting the conversion efficiency to be~\citep{budden}
\beq
\mathcal{C}^{\rm{(Budden)}}=\mathcal{T}(1-\mathcal{T}).
\label{cbudden}
\eeq
If the transmitted FW undergoes  additional wave reflection,
once a~right-hand polarized \mbox{${\rm R}$-cutoff} $(\epsR=\npar^2)$ or a supplementary MC layer is present in a~plasma,
the resulting reflection and mode conversion coefficients depend on the
constructive/destructive interference
between the partially reflected fast waves~\citep{majeski, fuchs, ram, kazakov2010, kim2011}.
For the cutoff-resonance-cutoff triplet (see Fig.~\ref{fig1}(b)), an accurate analytical
expression for the conversion efficiency was derived by \citet{fuchs} and \citet{ram}. It is given by
\beq
\mathcal{C}^{\rm{(triplet)}}=4\mathcal{T}(1-\mathcal{T})\sin^2(\Delta \phi/2).
\label{ctrip}
\eeq
Whereas for the isolated MC layer, the maximum MC efficiency is only 25\%, for the triplet
configuration it may be enhanced up to 100\%.
The enhanced MC efficiency requires the following two
conditions to be satisfied simultaneously: (i)~the phase difference is $\Delta \phi = (2n+1)\pi, n \in \mathbb{Z}$
(the reflected waves have
opposite phases and tend to cancel each other,
thereby minimizing the resultant
reflected wave); (ii) the MC layer is semi-transparent $\eta = \ln2/\pi \approx
0.22$ $(\mathcal{T} = 1/2)$. The latter condition for maximizing MC efficiency
is common for the Budden and triplet models, and our analysis focuses on identifying
proper conditions for that.

\vspace{-0.3mm}

\citet{prl2013} derived an analytical expression for the tunneling
factor for FW with $\kpar \simeq 0$ as a function of the heavy ion
concentration. If the heavy ion concentration is not too small,
the tunneling factor for plasmas of the terrestrial and the
Hermian magnetospheres was proven to be $\eta \gg 1$.  Provided the MC layer is non-transparent $(\eta \gg 1)$,
both Budden and triplet models predict that the incoming wave is almost totally
reflected at L-cutoff and can not tunnel via the barrier and undergo mode conversion.
E.g., for $\eta = 2$,
$\mathcal{C}^{\rm (Budden)} \approx 0.2\%$, whereas
$\mathcal{C}^{\rm {(triplet)}}_{\max} = 4\mathcal{T}(1-\mathcal{T}) \approx 0.8\%$.
On the contrary, it was shown that MC can be efficient for waves with
$\kpar/\kpar^{*} \lesssim 1$, where $\kpar^{*}$ is the FW critical
parallel wave number defined as
\beq
\kpar^{*} = \frac{\omega_{\rm pH}}{c} \frac{(f/f_{\rm cH})}{\sqrt{\zn_1 + \zn_2}}.
\label{kcrit}
\eeq
Here, $\omega_{\rm pH} = \sqrt{4 \pi n_e e^2/m_{\rm H}}$ is a~reference
proton plasma frequency, $\zn_1=Z_1/A_1=1$ and $\zn_2=Z_2/A_2$
are the ratios of the charge number to the atomic mass for protons and heavy
ions, respectively, and $f/f_{\rm cH}$ is the ratio of the measured resonant
ULF frequency to the local gyrofrequency of protons. Note that $\kpar^{*}$ depends on the plasma
density and scales linearly with the wave frequency and is independent
of the heavy ion concentration and the radius of the planet.  When
$\kpar$ is close to the critical $\kpar^*$, the FW dispersion includes
${\rm R}$-cutoff \citep{kim2014} and evaluation of $\eta$ from Eq.~(\ref{eq:eta}) allows to estimate
the tunneling of the incoming FW and determine whether MC can be efficient~\citep{prl2014}.

\section{Inferring heavy ion concentration}
\label{sect:3}

Typically, the concentration of heavy ions in magnetospheric plasmas
is not known \emph{a~priori}. Here, we describe briefly how this can be
estimated from the recorded resonant ULF spectrogram if the assumption of two-ion species plasma
is justified.
As follows from
Eqs.~(\ref{eq:tensor.comps}) and (\ref{eq:IIH}), the position of the
IIH resonance in a plasma depends mainly on three quantities: wave
frequency, heavy ion concentration and FW parallel wave number.  For
the measured wave frequency $f=f_{\rm obs}$, a~set of
$(\kpar,X_2)$--values satisfies the FW resonant condition
$\epsilon_{1} = n_{\|}^2$
at the given observation point
(cf. Figs.~\ref{fig1}(a) and (b) computed for different $X[{\rm He}^{+}]$)
\begin{equation}
\begin{array}{l}
\displaystyle
X_2= X_{\min} + (X_{\max} - X_{\min})(1 - (\kpar/\kpar^*)^2), \\[6pt]
\displaystyle
X_{\min} = \frac{(f/f_{\rm cH})^2 - \zn_2^2}{\zn_1^2 - \zn_2^2},
X_{\max} = \frac{\mu X_{\min}}{1+(\mu-1)X_{\min}},
\label{xvskpar}
\end{array}
\end{equation}
where $X_2 = Z_2 n_2/n_e$ is the \emph{unknown} concentration of heavy ions multiplied by their charge number,
and $\mu = \zn_1/\zn_2$.
As an \emph{input} value for inferring the heavy ion concentration, we use
the ratio of the resonant ULF frequency to the local gyrofrequency of protons
$f/f_{\rm cH}$. If one calculates the exact value of $\kpar$ yielding the maximized MC,
the~corresponding heavy ion concentration can be determined from Eq.~(\ref{xvskpar}).
Its value is between $X_{\min}$ and $X_{\max}$, both of which, in turn, depend on $f/f_{\rm cH}$.

As an example, we consider two linearly polarized ULF wave events,
at Mercury's magnetosphere (including a significant fraction of heavy
sodium ions $\Na^{+}$) detected by the Mariner 10 spacecraft
and at the Earth's magnetosphere (including helium ions) detected by the GEOS satellite.
In the case of Mercury, the spectral intensity was peaked at frequency $f \approx
0.5~{\rm Hz}$~\citep{russell}.  The local magnetic field strength was
\mbox{$B \approx 86~{\rm nT}$} and hence proton and sodium gyrofrequencies
were $f_{\rm cH} \approx 1.31~{\rm Hz}$ and $f_{\rm c,{\rm Na}^{+}}
\approx 0.057~{\rm Hz}$. In the case of Earth, the observed frequency
was $f \approx 1.0~{\rm Hz}$, whereas the local proton gyrofrequency
was $f_{\rm cH} \approx 2.28~{\rm Hz}$ \citep{perraut}.

\begin{figure}
\centering
\includegraphics[trim=0.cm 0cm -0cm -0mm, clip=true,
height=0.37\textwidth]{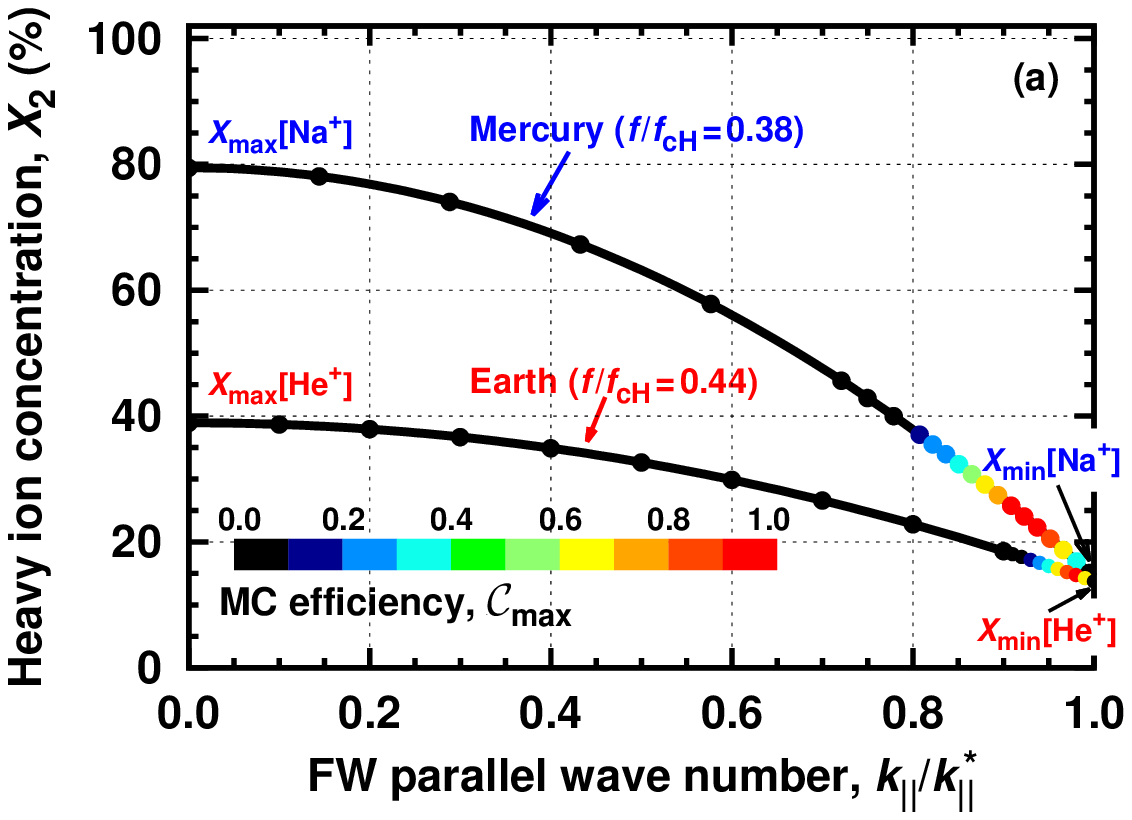}
%\hfill
\includegraphics[trim=0cm 0cm -0mm -0.0cm, clip=true,
height=0.37\textwidth]{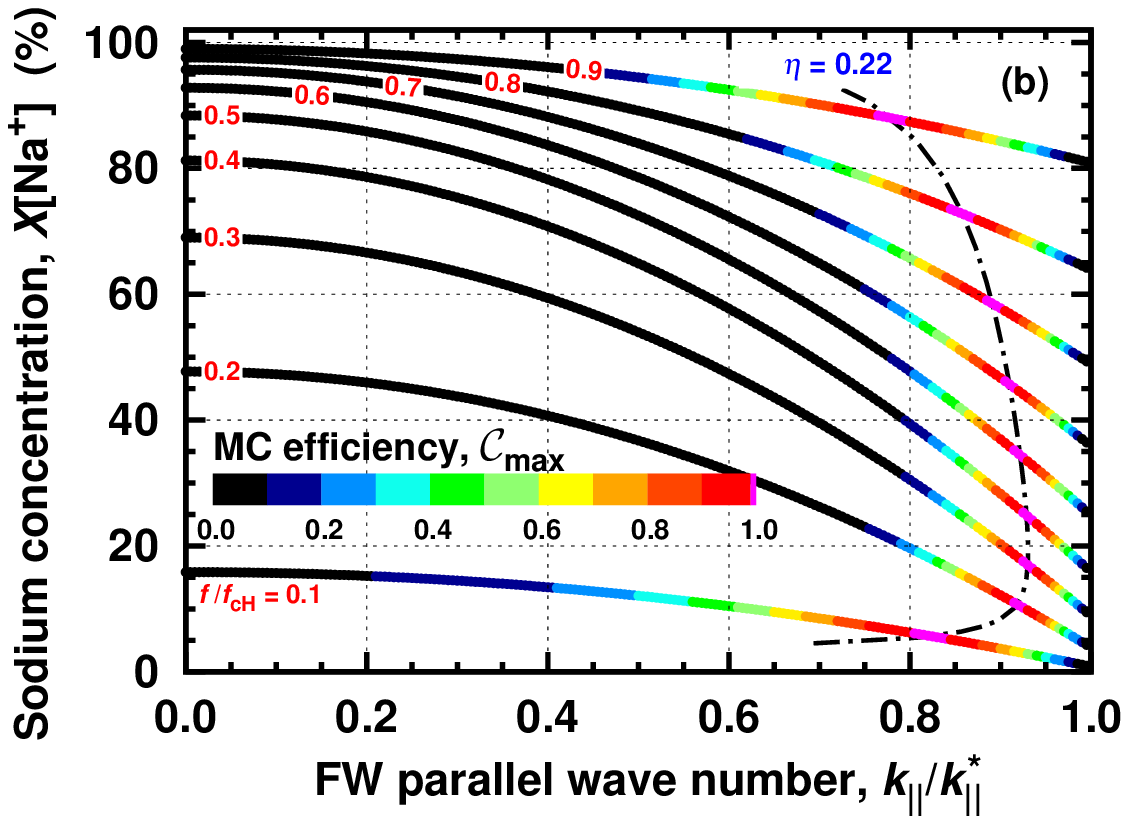}
\vspace{-5mm}
\caption{ (a) Heavy ion concentration vs. the FW parallel wave number for the fixed IIH resonance
location.
The colour of the points represents the maximum MC efficiency,
\mbox{$\mathcal{C}_{\max} = 4\mathcal{T}(1 - \mathcal{T})$}.
(b) The $(k_{\|},X[{\rm Na}^{+}])$--diagram for the different $f/f_{\rm cH}$ ratios for Mercury's plasma.
The dash-dotted line shows $\eta=0.22$ and maximized MC efficiency ($\mathcal{C}_{\max}=100\%$).
\label{fig2}
}
\end{figure}

Figure~\ref{fig2}(a)~shows the $(\kpar,X_2)$--diagram computed for the
conditions of Earth's ($R=6.6R_{\rm E}$ and $n_{e}=10~{\rm cm}^{-3}$)
and Mercury's ($R=1.5R_{\rm M}$ and $n_{e}=3~{\rm cm}^{-3}$)
magnetospheres, choosing the input frequency ratios $f/f_{\rm cH}
\approx 0.44$ and $0.38$, respectively.
For the chosen $f/f_{\rm cH}$ and for any individual $\kpar$,
the dispersion relation (\ref{eq:fw.disp}) is evaluated numerically
for different $X_2$ finding the proper one (shown with circles), for which
the IIH resonance is located at the considered radial position.
The solid lines show the heavy ion concentration as
predicted by Eq.~(\ref{xvskpar}).
It is clear that there is excellent agreement
between the numerical results and Eq.~(\ref{xvskpar}), and the algebra behind this equation is correct.
In addition, for every pair of $(\kpar,X_2)$ values we evaluated numerically the
tunneling factor using Eq.~(\ref{eq:eta}) and the maximum MC efficiency
\mbox{$\mathcal{C}_{\max}=4\mathcal{T}(1-\mathcal{T})$}.  The latter
is represented by the colour of the circles.

\vspace{-0.3mm}

Taking the Mercury case as an example, one can consider two different
extreme cases for the measured $f/f_{\rm cH} \approx 0.38$.  If the
detection of waves with $\kpar \simeq 0$ is assumed, then the IIH
resonant frequency $\omega_{\rm S}$ reduces to the well-known Buchsbaum
two-ion hybrid frequency
$$
\omega_{\rm S0} = \omega_{\rm c2}
\sqrt{\frac{1-(1-\mu)X_2}{1-(1-1/\mu)X_2}},
$$
which satisfies $\epsilon_1(\omega_{\rm S0})=0$. However, for such $\kpar$ and for the considered
plasma conditions of Earth and Mercury, $\eta \gg 1$ and the MC layer is
non-transparent, and therefore the FW cannot tunnel
through the barrier and undergo mode conversion~\citep{prl2013}.
The other limiting case corresponds to $\kpar = \kpar^{*}$ and $X_2 =
X_{\min}$. For these parameters, the IIH resonance, L-cutoff and
R-cutoff intersect at the observation point such that the MC layer
width and accordingly also the tunneling factor reduce to zero.
This occurs for $\epsilon_2(\omega_{\rm cross}) = 0$
($\eps_2$ is a non-diagonal tensor element),
and is typically referred to
as the crossover condition~\citep{othmer}.  The crossover frequency in
a two-ion species plasmas is linked to the heavy ion concentration as
$$
\omega_{\rm cross} = \omega_{\rm c2} \sqrt{1 + (\mu^2 - 1)X_2}.
$$

According to Fig.~\ref{fig2}(a), for the measured frequency ratio for Mercury's
plasma ($f/f_{\rm cH} \approx 0.38$), the sodium concentration is
between $X_{\min}[\Na^{+}] = 14.3\%$ and $X_{\max}[\Na^{+}] = 79.3\%$.
For the frequency ratio $f/f_{\rm cH} \approx 0.44$ reported for Earth
by~\citet{perraut},  Eq.~(\ref{xvskpar}) implies that the helium
concentration is between $X_{\min}[{\rm He}^{+}] =$ $14.0\%$ and
$X_{\max}[{\rm He}^{+}] = 39.4\%$.  These values, which correspond to the
Buchsbaum ($\kpar = 0$, $\eta \gg 1$) and the crossover ($\kpar =
\kpar^{*}$, $\eta =0$) conditions, yield the \emph{upper} and the
\emph{lower} estimates for the heavy ion concentration, respectively.

Since efficient MC requires $0.05 \leq \eta \leq 0.61$ (these are the tunneling
factors at which $\mathcal{C}_{\max}=50\%$) and is maximized at
$\eta \approx 0.22$ \citep{prl2013}, one
can conclude that the FW parallel wave number should be close to, but
somewhat below, $\kpar^{*}$. This is also illustrated in
Fig.~\ref{fig2}(b), which shows the $(\kpar,X_2)$--diagram for various
$f/f_{\rm cH}$ in the case of Mercury.  The dash-dotted line in
Fig.~\ref{fig2}(b) depicts the optimal case $\eta = 0.22$, which leads to
the maximized MC efficiency $(\mathcal{C}_{\max}=100\%)$
and determines the sodium concentration as a function of the observed frequency.

For $X_2 = X_{\min}$, decreasing $\kpar$ will result in a non-zero MC
layer width.  Simultaneously, the resonant layer will be shifted to
the higher magnetic field side relative to the observation point.  As
a~result, the heavy ion concentration should be somewhat larger than $X_{\min}$
to displace the resonant
layer in the opposite direction and locate it back at the observation
point (for the fixed wave frequency).  This opposite shift of the IIH
resonant layer due to the change of $\kpar$ and $X_2$ is captured by
Eq.~(\ref{xvskpar}): for any heavy ion concentration within the range
$X_{\min} \leq X_2 \leq X_{\max}$, there exists a~particular wave
number $\kpar$, for which the IIH resonance is located at the point of
observation.

\begin{figure}
\centering
\includegraphics[trim=0cm 0cm -0mm -0.0cm, clip=true,
height=0.37\textwidth]{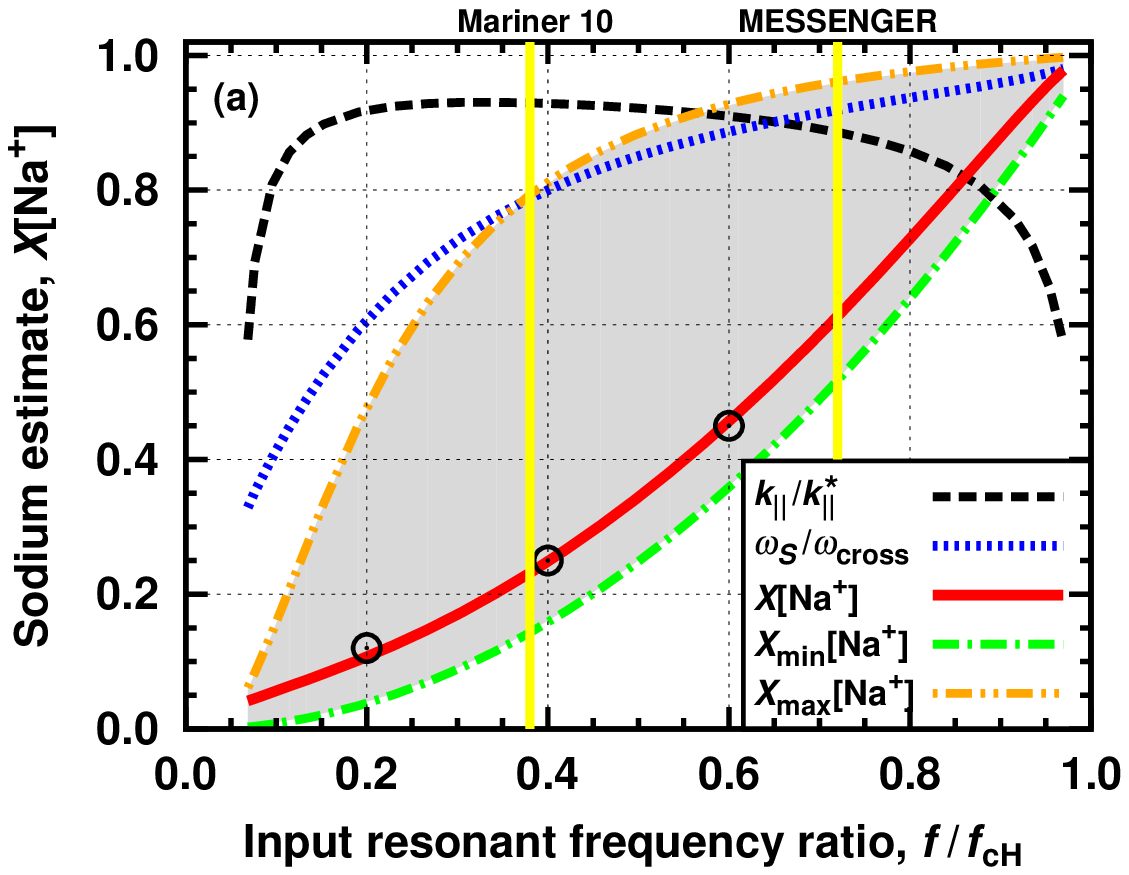}
\includegraphics[trim=0cm 0cm -0mm -0.0cm, clip=true,
height=0.37\textwidth]{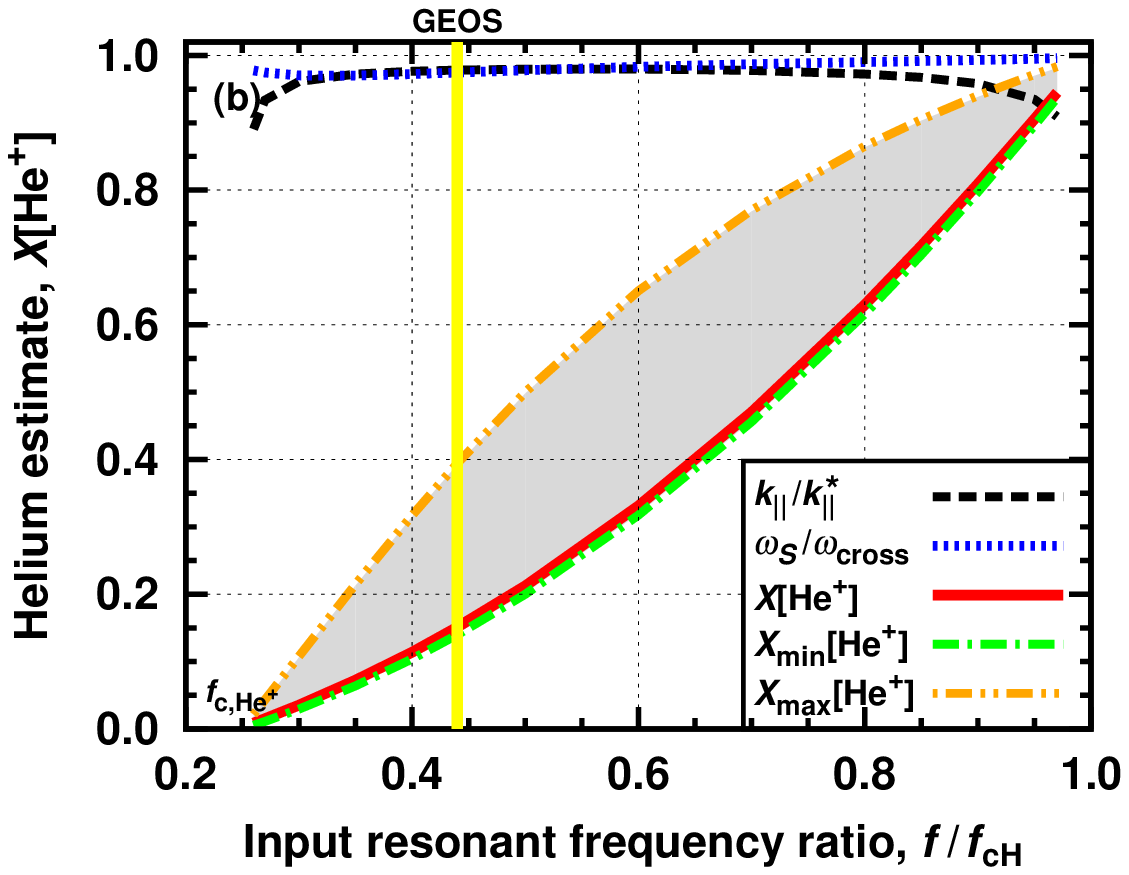}\\
{
\caption{ (a) and (b) Sodium and helium concentrations vs. the
measured frequency ratio~$f/f_{\rm cH}$ for the magnetospheres
of Mercury and Earth, respectively. Open circles in Fig.~\ref{fig3}(a)
correspond to the data reported by \citet{kim2011}.
\label{fig3}
}}
\end{figure}

As follows from Fig.~\ref{fig2}(a), the maximum MC efficiency
$(\mathcal{C}_{\max}=1)$ is reached at \mbox{$\kpar/\kpar^{*} \approx 0.93$}
and $0.98$ for the plasmas of Mercury and Earth, respectively. Once
this wave number ratio is known, the heavy ion concentration can be
easily evaluated.  For Earth, Eq.~(\ref{xvskpar}) yields \mbox{$X[{\rm He}^{+}] \approx 15.1\%$},
while for Mercury $X[{\rm Na}^{+}] \approx 23.3\%$.
The computed value for the sodium concentration -- as expected -- 
is somewhat larger than $X[{\rm Na}^{+}] \approx 14\%$ reported by \citet{othmer},
who analyzed the same wave event and used the crossover frequency itself as a preferred resonant ULF frequency.
Note that the values we obtain for $X[{\rm Na}^{+}]$ are in very good
agreement with the results of more sophisticated numerical modeling
performed by \citet{kim2011}, who accounted for the additional FW reflection
at R-cutoff and wave interference (shown with open circles in Fig.~\ref{fig3}(a)).
For $\omega_{\rm S}/\omega_{\rm cH}=0.4$,
a~sodium concentration $X[{\rm Na}^{+}]=25\%$ was reported. This is not surprising since -- as already
mentioned -- efficient mode conversion requires the evanescence layer to be
a semi-transparent $(\mathcal{T} \simeq 1/2)$, regardless if the Budden or the triplet MC model is used.
Figure~\ref{fig2}(b) is also consistent with another two cases considered by~\citet{kim2011},
viz. $(\omega_{\rm S}/\omega_{\rm cH};X[\Na^{+}])=(0.2;12\%)$ and $(0.6;45\%)$.

The ratio of the IIH resonant to the crossover frequency is also
linked to the $\kpar/\kpar^{*}$ value
\begin{equation}
\begin{array}{l}
\displaystyle
p = \omega_{\rm S}/\omega_{\rm cross} = \left[ 1 + G(x) \left( 1 - (\kpar/\kpar^{*})^2 \right) \right]^{-1/2}, \\[16pt]
\displaystyle
G(x) = \frac{(x^2 - \zn_2^2)(\zn_1^2 - x^2)}{x^2(\zn_1 \zn_2 + x^2)},
\label{eq1}
\end{array}
\label{pratio}
\end{equation}
where we introduced $x=f/f_{\rm cH}$. Equation~(\ref{pratio}) predicts $p \approx 0.97$ for
the plasma conditions of the terrestrial magnetosphere near the geostationary orbit,
whereas for Mercury's plasma $p \approx 0.79$, in agreement with the values reported by~\citet{prl2014}.
Vice versa, if the $\omega_{\rm S}/\omega_{\rm cross}$ ratio is known,
by combining Eqs.~(\ref{xvskpar}) and (\ref{pratio}), the next-order
correction to the heavy ion concentration can be determined to be
\beq
X_2/X_{\min} = (x^2/p^2 - \zn_2^2)/(x^2 - \zn_2^2) \gtrsim 1.
\eeq
For Mercury, this can be simplified further since $\zn_2=1/23 \ll 1$, and
$X[{\rm Na}^{+}] \approx X_{\min}[{\Na}^{+}]/p^2$.

As provided by our simplified MC analysis, Figure~\ref{fig3}
illustrates the estimates of the sodium and helium concentrations as functions of
resonant frequency in the magnetospheres of Mercury and Earth,
respectively.
With green dash-dotted and orange dash-double-dotted
lines we show the minimum and maximum values of the heavy ion
concentrations {(Eq.~(\ref{xvskpar}))},
and with red solid line we show the estimated heavy ion
concentration {determined from $\mathcal{C}_{\max}=100\%$}.
In the same figure, the ratio
$\kpar/\kpar^{*}$ and $\omega_{\rm S}/\omega_{\rm cross}$
is plotted with dashed and dotted line, respectively.
Note~that whereas for Mercury's
plasma the resonant frequency $\omega_{\rm S}$ is close
to the crossover frequency
{(according to Fig.~\ref{fig3}(a),
$p \approx 0.8$ for Mariner 10 and $p \approx 0.9$ for MESSENGER observations,
$f/f_{\rm cH} \simeq 0.7$ reported by \citet{boardsen})},
for~the magnetosphere of Earth the resonant IIH
frequency is even closer to the crossover frequency $(p \approx 0.97) $
(see Fig.~\ref{fig3}(b)).

\vspace{-0.3mm}

The numerical results shown in Fig.~\ref{fig3} can be interpreted by
using Eq.~(\ref{pratio}). As~a~function of $f/f_{\rm cH}$, $G(x)$ in
Eq.~(\ref{pratio}) maximizes at $\hat{x} =
\zn_1/\sqrt{\mu^{3/2}-\mu+\mu^{1/2}}$ and equals to
$\hat{G}=(1+1/\mu)(\sqrt{\mu}-1)^2$. This yields for Earth
$\hat{G}_{\rm Earth}=1.25$ but it is substantially larger for Mercury
$\hat{G}_{\rm Mercury} \approx 15.0$. For the measured $f/f_{\rm cH}$
values, the magnitude of the $G$--term and the difference in the
computed $\kpar/\kpar^{*}$ values, contribute equally to the fact why
the $\omega_{\rm S}/\omega_{\rm cross}$ ratio is much closer to one for
Earth than for Mercury ($G_{\rm Earth}(0.44) \approx 1.2$ vs. $G_{\rm
  Mercury}(0.38) \approx 4.5$).

The fact that for plasmas near the geostationary orbit,
the resonant IIH frequency is very close to the crossover frequency, potentially
allows an accurate determination of $X[{\rm He}^{+}]$ by matching
$f_{\rm obs}$ to $f_{\rm cross}$, without a~need to go to the
second-order approximation as in case of Mercury.  However, this
holds only if the two-ion species approximation is justified.
Otherwise, one needs to account for the upshift/downshift of the
crossover frequency due to the presence of additional ion species, e.g. O$^+$.
This effect is more important for the Earth's magnetosphere than the
correction due to the finite $p$--value.
Following the reasoning presented in \citet{prl2013},
one can show that the heavy ion concentration satisfying the crossover
condition, in general case, is given by
%%%%%%%%%%%%%%%%%%%%%%%%%%%%%%%%%%%%%%%%%%%%%%%%%%%%%%%%%%%%%%%%%%%%%%
%
\beq
\sum_{i=2,3,...} \frac{\zn_1^2 - \zn_{i}^2}{(f/f_{\rm cH})^2 - \zn_{i}^2}\,X_{i} = 1,
\label{eq:multiion}
\eeq
%
%%%%%%%%%%%%%%%%%%%%%%%%%%%%%%%%%%%%%%%%%%%%%%%%%%%%%%%%%%%%%%%%%%%%%%
%
where the sum is over all ion species, except of the majority proton ions.
Similar analysis was given by \citet{othmer} to correct the
sodium estimate accounting for additional contamination of Hermean plasma
with He$^{2+}$ ions. If we use the same frequency ratio $f/f_{\rm obs}=0.38$ reported
by \citet{russell},
Eq.~(\ref{eq:multiion}) predicts
$X[{\rm Na^+}] \approx 0.143 + 2.03 n_{{\rm He}^{2+}}/n_{e}$
(note that in our notations $X[{\rm He}^{2+}]=2n_{{\rm He}^{2+}}/n_{e}$).
Hence, Eq.~(\ref{eq:multiion}) is the
generalization of Eq. (33) by \citet{othmer}.
Note that Eq.~(\ref{eq:multiion}) is not restricted for plasmas with three ion species
but is correct for an arbitrary number of ion species.
However, a complete determination of the plasma composition
is not possible in this case (the number
of unknowns exceeds the number of equations) without having  information on
the concentrations of additional ion species or setting up another
constraints. For example, for Mercury's plasma \citet{othmer} assumed
a mixing ratio for Na$^{+}$ and He$^{2+}$ constituents to be the same as in the
the  solar  wind, and that provided the additional constraint necessary for determining the plasma composition.

\begin{figure}
\centering
\includegraphics[trim=0cm 0cm -0mm -0.0cm, clip=true, height=0.5\textwidth]{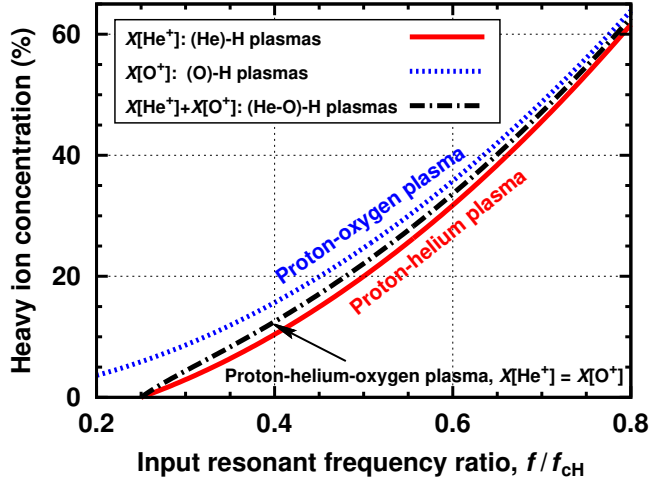}
\caption{Heavy ion concentrations satisfying the crossover frequency in~proton-helium-oxygen plasmas.}
\label{fig4}
\end{figure}

Figure~\ref{fig4} shows the heavy ion concentrations satisfying the crossover
frequency in proton-helium-oxygen plasmas.
The red solid curve for the proton-helium plasma is exactly the same
as the curve showing $X_{\min}[{\rm He}^{+}]$ in Fig.~\ref{fig3}(b).
The blue dotted line depicts another limiting case of
the proton-oxygen plasma. For the same resonant frequency ratio,
the estimated oxygen concentration is somewhat higher than that of helium,
in line with Eq.~(\ref{xvskpar}).
For $f/f_{\rm cH} = 0.44$, Eq.~(\ref{eq:multiion}) predicts $X[{\rm He}^{+}] + 0.73X[{\rm O}^{+}] \approx 0.14$,
and one gets $X[{\rm He}^{+}]=14\%$ for (He)-H plasmas and $X[{\rm O}^{+}]=19\%$ for (O)-H plasmas.
Figure~\ref{fig4} also displays the intermediate case, when the helium and oxygen concentrations
are assumed to be equal (dashed-dotted line).
For the given resonant ULF frequency, $X[{\rm He}^{+}] + X[{\rm O}^{+}] \approx 16\%$.
This example shows that the presented approach still provides reasonably fair estimate
for the total heavy ion concentration, even if one has no a~priori information
on the helium/oxygen density ratio.

\vspace{-0.3mm}

\section{Tunneling factor at the crossover condition}
\label{sect:4}
\begin{figure}
\centering
\includegraphics[trim=0cm -0.0cm -0mm -0.0cm, clip=true,
height=0.37\textwidth]{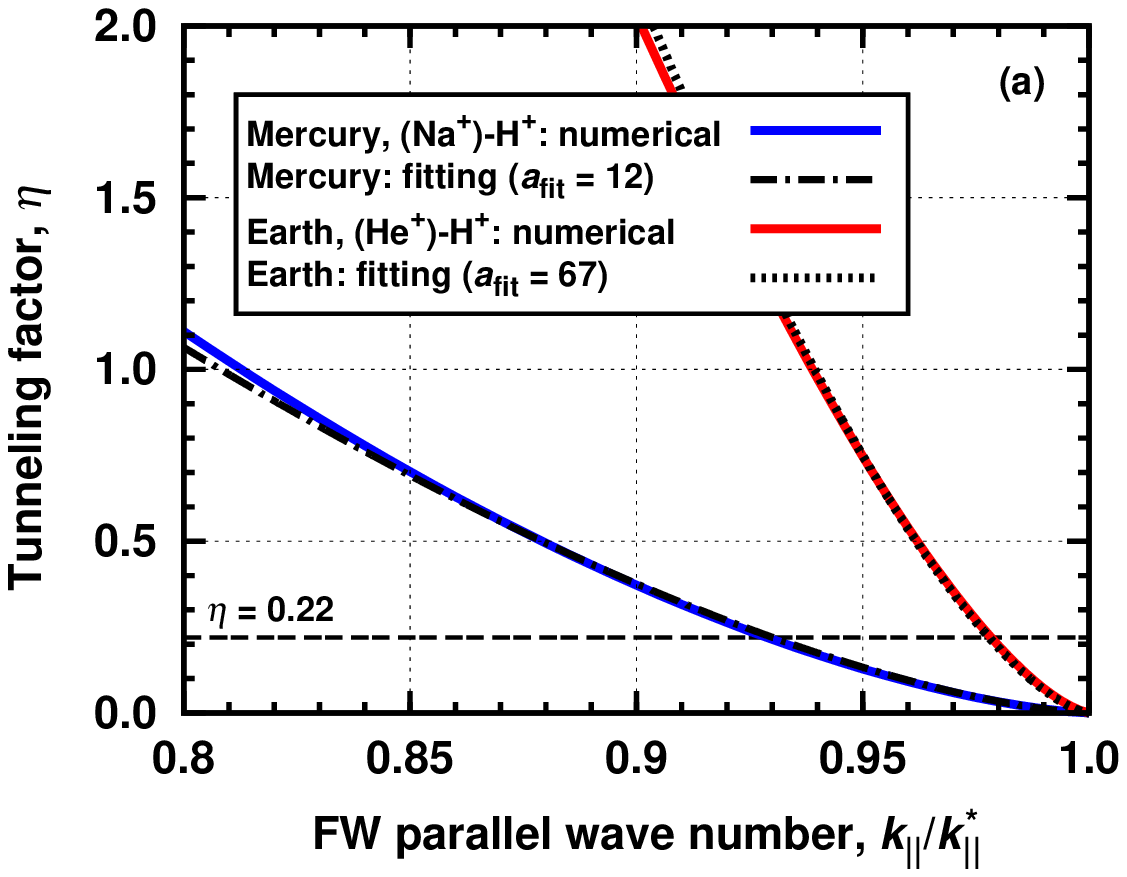}
\includegraphics[trim=0cm -0.0cm -0mm -0.0cm, clip=true,
height=0.37\textwidth]{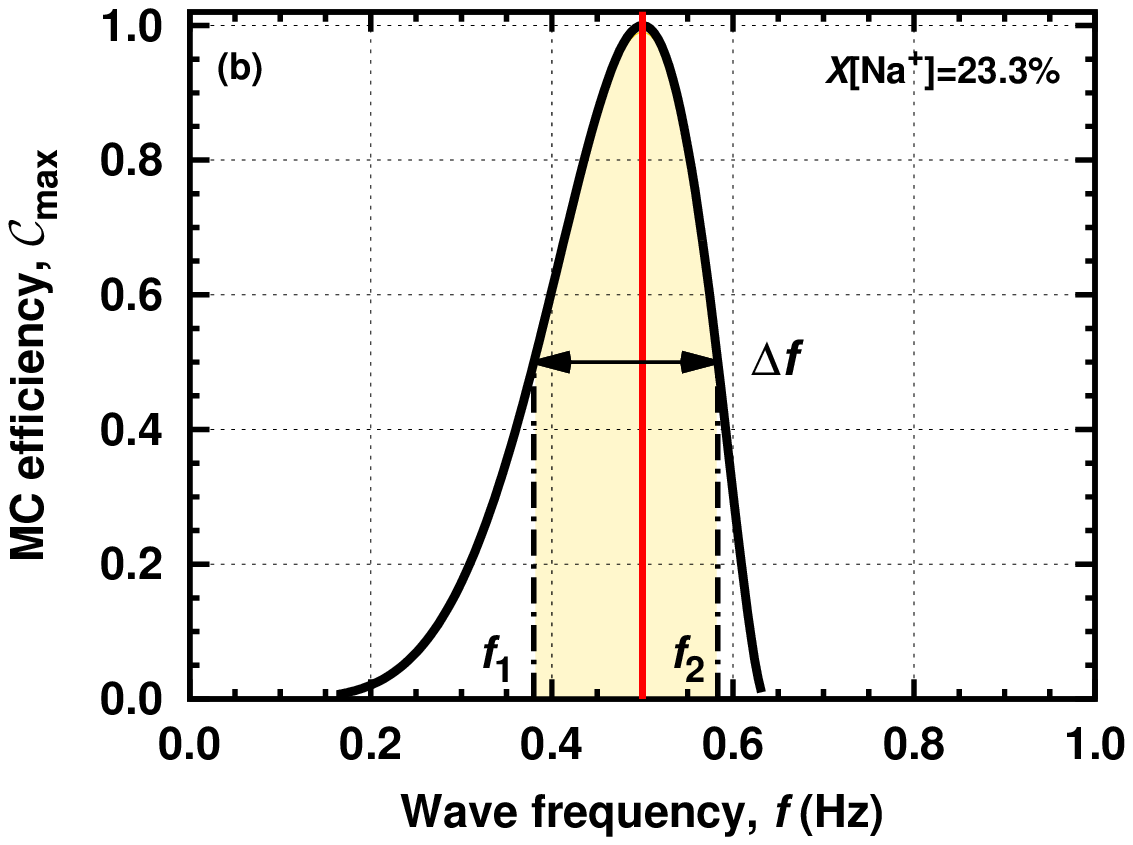}
{
\caption{(a) Tunneling factor vs. $\kpar$ at the crossover condition.
(b) Computed MC efficiency $\mathcal{C}_{\max}$ vs. the wave frequency
for Mercury's plasma and $X[\Na^{+}] = 23.3\%$.
}
\label{fig5}
}
\end{figure}

By applying the least-square fitting of the numerical results for
$\eta$ computed using Eq.~(\ref{eq:eta}), we find that the tunneling factor for $\kpar/\kpar^{*}
\lesssim 1$ scales as (see Fig.~\ref{fig5}(a))
\beq
\eta(\kpar) \approx a_{\rm fit} \left( 1 - \kpar/\kpar^{*} \right)^{3/2}.
\label{eq:eta.approx}
\eeq
For the considered conditions of magnetospheric plasmas, $a_{\rm fit}
\approx 12$ for Mercury and $a_{\rm fit} \approx 65$ for Earth (assuming (He)-H plasma),
respectively.
Then, using Eq.~(\ref{eq:eta.approx}), the FW parallel
wave number, at which $\eta = 0.22$ and maximized MC occurs, can be
derived
\beq
k_{\|}^{\dagger}/\kpar^{*} = 1 - 0.365 a_{\rm fit}^{-2/3}.
\eeq
If we use the computed numerical values for $a_{\rm fit}$,
$k^{\dagger}_{\|}/\kpar^{*} \approx 0.93$ and $0.98$ are obtained,
consistent with previous results. This approach also allows to
estimate the range of $\kpar$, under which an efficient MC occurs
$(0.05 \leq \eta \leq 0.61$, i.e. $\mathcal{C}_{\max} \geq 50\%)$ once
only a single value of
$k_{\|}^{\dagger}$ is known
\beq
\Delta \kpar/\kpar^{*} \approx 1.6(1 -
k^{\dagger}_{\|}/\kpar^{*}).
\label{1.6}
\eeq
The coefficient, which appears in
the right-hand side of Eq.~(\ref{1.6}), reflects the power exponent identified
in the scaling, viz.
$1.6 \approx (0.61^{2/3} - 0.05^{2/3})/(0.22^{2/3})$.
%%%%%%%%%%%%%%%%%%%%%%%%%%%%%%%%%%%%%%%%%%%%%%%%%%%%%%%%%%%%%%%%%%%%%%%%%%%%%
%
%   1.6 = (1 + ln(2+sqrt2)/ln2)^{2/3} - (1 + ln(2-sqrt2)/ln2)^{2/3}
%
%   0.22=ln2/pi;    0.05=ln(2(2-sqrt2))/pi;    0.61=ln(2(2+sqrt2))/pi
%
%%%%%%%%%%%%%%%%%%%%%%%%%%%%%%%%%%%%%%%%%%%%%%%%%%%%%%%%%%%%%%%%%%%%%%%%%%%%%
This means that for Mercury's plasma the
$\kpar$--spectrum where efficient MC occurs has a width $\Delta \kpar/\kpar^{*}
\approx 0.11$, but for the Earth's plasma it is narrower, $\Delta
\kpar/\kpar^{*} \approx 0.03$.
Note that since our approach is based on evaluating $\mathcal{C}_{\max}$
and neglecting the wave interference and the phase term in Eq.~(\ref{ctrip}), the
presented approximation for $\Delta k_{\|}$ overestimates the actual value.

\vspace{0mm}

\section{Frequency width of the wave spectrum}
\label{sect:5}

The measured wave spectrum has a finite frequency width $\Delta f$.
For the ULF event in Mercury's magnetospheric plasma
$\Delta f \approx 0.26-0.30~{\rm Hz}$ was reported by \citet{russell}.
Here, we investigate
if that is consistent with our analysis and the assumption that
the observed resonant ULF waves are due to mode conversion.
To accomplish that, we apply
a somewhat different reasoning: the sodium concentration is kept
constant $(X[{\rm Na}^{+}] \approx 23.3\%)$, and the parallel wave
number $\kpar$ is adjusted for every frequency $f$ according to
Eq.~(\ref{xvskpar}), such that the IIH resonance is located at the point
of observation.  Figure~\ref{fig5}(b) shows the computed
$\mathcal{C}_{\max}$ as a function of $f$.  Again, the Buchsbaum
$(\kpar=0, \eta \gg 1)$ and the crossover conditions
$(\kpar=\kpar^{*}, \eta = 0)$ can be clearly distinguished, represented by the
two side frequencies at which $\mathcal{C}_{\max} \rightarrow 0$. In terms of
the wave frequency these can written as
\begin{equation}
\begin{array}{l}
\displaystyle
f_{\min} = f_{\rm c2} \sqrt{\frac{1-(1-\mu)X_2}{1-(1-1/\mu)X_2}} \approx 0.16~{\rm Hz}, \\[16pt]
\displaystyle
f_{\max} = f_{\rm c2} \sqrt{1 - (1- \mu^2)X_2} \, \approx 0.635~{\rm Hz},
\label{eq11}
\end{array}
\end{equation}
respectively. If we assume that the measured spectral intensity of the resonant linearly polarized ULF signal
is proportional to the MC efficiency,
then $\Delta f$ requires the evaluation of the full width at half maximum for $\mathcal{C}_{\max}$:
$\mathcal{C}_{\max} = 50\%$ is
reached at $f_{1} \approx 0.38~{\rm Hz}$
and $f_{2} \approx 0.58~{\rm Hz}$.
Hence, for the considered conditions the spectrum
is peaked at $\hat{f}=0.5~{\rm Hz}$ and has a frequency width
$\Delta f = f_2 - f_1 \approx 0.20~{\rm Hz}$.
The computed value is in a qualitative agreement with the observations.

\section{Conclusions}
\label{sect:6}
We have complemented previous analytical results derived in
\citep{prl2013} by evaluating the tunneling factor and the
maximum MC efficiency $\mathcal{C}_{\max}$  numerically for the magnetospheres of
Earth and of Mercury. Efficient MC is shown to occur in a~narrow range
of $\kpar$ close to the FW critical wave number $\kpar^{*}$, which
depends mainly on the plasma density and the frequency $f/f_{\rm cH}$ ratio, and
is independent of the radius of the planet and the heavy ion
concentration.  For the detected wave event in Mercury's
magnetosphere, $\omega_{\rm S}$ is confirmed to be close, but somewhat
below, the crossover frequency {($p \approx 0.8$ for Mariner 10 and $p \approx 0.9$ for MESSENGER
observations)}.  For the plasma conditions
near the geostationary orbit, the resonant IIH frequency is
very close to the crossover frequency for
a~wide range of parameters $(p \approx 0.97)$, potentially making an
accurate estimate of the heavy ion concentration from some of the detected resonant
linearly polarized ULF signals possible.  For Mercury, the evaluated
frequency width of the spectrum $\Delta f$ has been shown to be
consistent with the measured value.  \\

\textbf{Acknowledgements.} The authors are grateful to the anonymous referees for the valuable improvements they suggested.

%\newpage
\bibliographystyle{jpp}
% Note the spaces between the initials
%\bibliography{jpp-instructions}

\end{document}